\documentclass[reprint,superscriptaddress,aps,prb,twocolumn]{revtex4-2}
\usepackage{setspace}
\usepackage{amsmath}
\usepackage{graphicx}
\usepackage{amsfonts}
\usepackage{amssymb}
\usepackage{epstopdf} 
\usepackage{xcolor}
\usepackage{verbatim}
\usepackage{soul}
\usepackage{comment}
\usepackage{placeins}
\usepackage{textcomp}
\usepackage{bm}
\usepackage[colorlinks=true,bookmarks=false,citecolor=blue,urlcolor=blue]{hyperref}

\DeclareGraphicsExtensions{.pdf,.eps,.png,.jpg,.mps} 

\usepackage{float}

\begin{document}

\title{
Brillouin Backaction Thermometry for Modal Temperature Control\\
}

\author{Yu-Hung Lai$^{1\ast}$, Zhiquan Yuan$^{1\ast}$, Myoung-Gyun Suh$^{1}$, Yu-Kun Lu$^{1}$, Heming Wang$^{1}$, Kerry J. Vahala$^{1\dagger}$ \\ 
$^1$T. J. Watson Laboratory of Applied Physics, California Institute of Technology, Pasadena, California 91125, USA \\
$^\ast$These authors contributed equally to this work. \\ 
$^\dagger$vahala@caltech.edu}

\begin{abstract}
Stimulated Brillouin scattering provides optical gain for efficient and narrow-linewidth lasers in high-Q microresonator systems. However, the thermal dependence of the Brillouin process, as well as the microresonator, impose strict temperature control requirements for long-term frequency-stable operation. Here, we study Brillouin back action and use it to both measure and phase-sensitively lock modal temperature to a reference temperature defined by the Brillouin phase-matching condition. At a specific lasing wavelength, the reference temperature can be precisely set by adjusting resonator free spectral range. This backaction control method is demonstrated in a chip-based Brillouin laser, but can be applied in all Brillouin laser platforms. It offers a new approach for frequency-stable operation of Brillouin lasers in atomic clock, frequency metrology, and gyroscope applications.
\end{abstract}

\maketitle

\begin{figure*}[t]
    \centering
    \includegraphics[width=\linewidth]{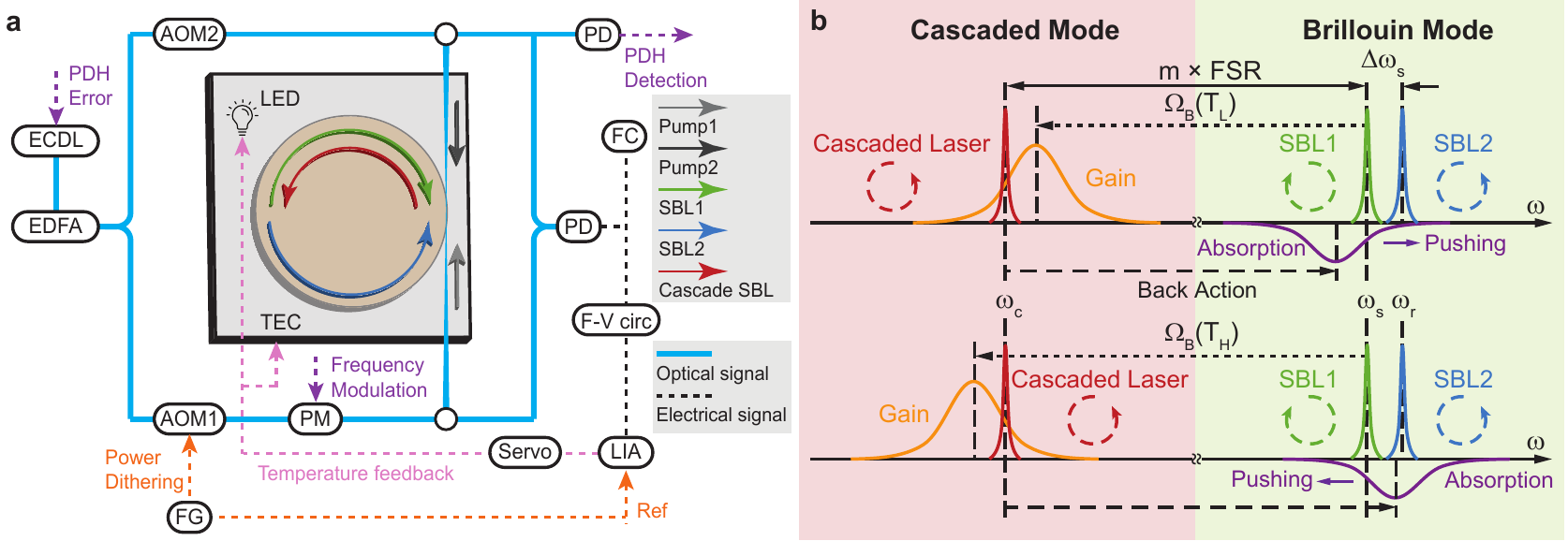}
    \caption{\textbf{Experimental setup and illustrations of the cascaded Brillouin laser induced backaction.} \textbf{a,} Pump 1 (counterclockwise, grey) and Pump 2 (clockwise, black) waves are generated from an external cavity diode laser (ECDL) amplified by an Erbium doped fiber amplifier (EDFA). Both the pump 1 and pump 2 frequencies are shifted using acousto-optic modulators (AOM) to create a relative frequency offset. The frequency of Pump 1 is further Pound-Drever-Hall (PDH) locked to the cavity resonance using a phase modulator (PM). Green (blue) arrow refers to Brillouin laser waves SBL1 (SBL2) discussed in panel {\bf b}. Pump 1 has a slightly higher power so that the cascaded SBL is generated in the counter-clockwise direction (red). The beat signal of SBL1 and SBL2 is generated by a photodetector (PD) and its frequency is measured by a frequency counter (FC) and then frequency discriminated using a frequency tracking circuit (F-V circ) for subsequent phase sensitive detection by a lock-in amplifier (LIA), and temperature control using a light emitting diode (LED) and thermal electrical cooler (TEC). 
    \textbf{b,} Upper panel (low temperature case, T$_{\rm L}$): the Brillouin shift $\Omega_B ({ \rm T}_{\rm L})$ is smaller than the mode frequency difference ($m\times$FSR, where $m$ the mode number difference and FSR is the free spectral range in angular frequency). Backaction on SBL1 pushes its frequency away from the backaction absorption (purple) maximum, thereby increasing its frequency. This decreases the SBL1-SBL2 beating frequency ($\Delta \omega_{s}/ 2 \pi$), when the cascaded laser power increases. Note: SBL2's frequency is not affected by the mode-pushing effect because the backaction absorption is directional according to the phase-matching condition. Lower panel (high temperature case, { \rm T}$_{\rm H}$): the Brillouin shift $\Omega_B ({\rm T}_{\rm H})$ is larger than the mode frequency difference so that SBL1 has its frequency pushed lower. This increases the SBL1-SBL2 beating frequency ($\Delta \omega_{s}/ 2 \pi$), when the cascaded laser power increases.}
\label{Fig1}
\end{figure*}

\begin{figure*}[t]
    \centering
    \includegraphics[width=\linewidth]{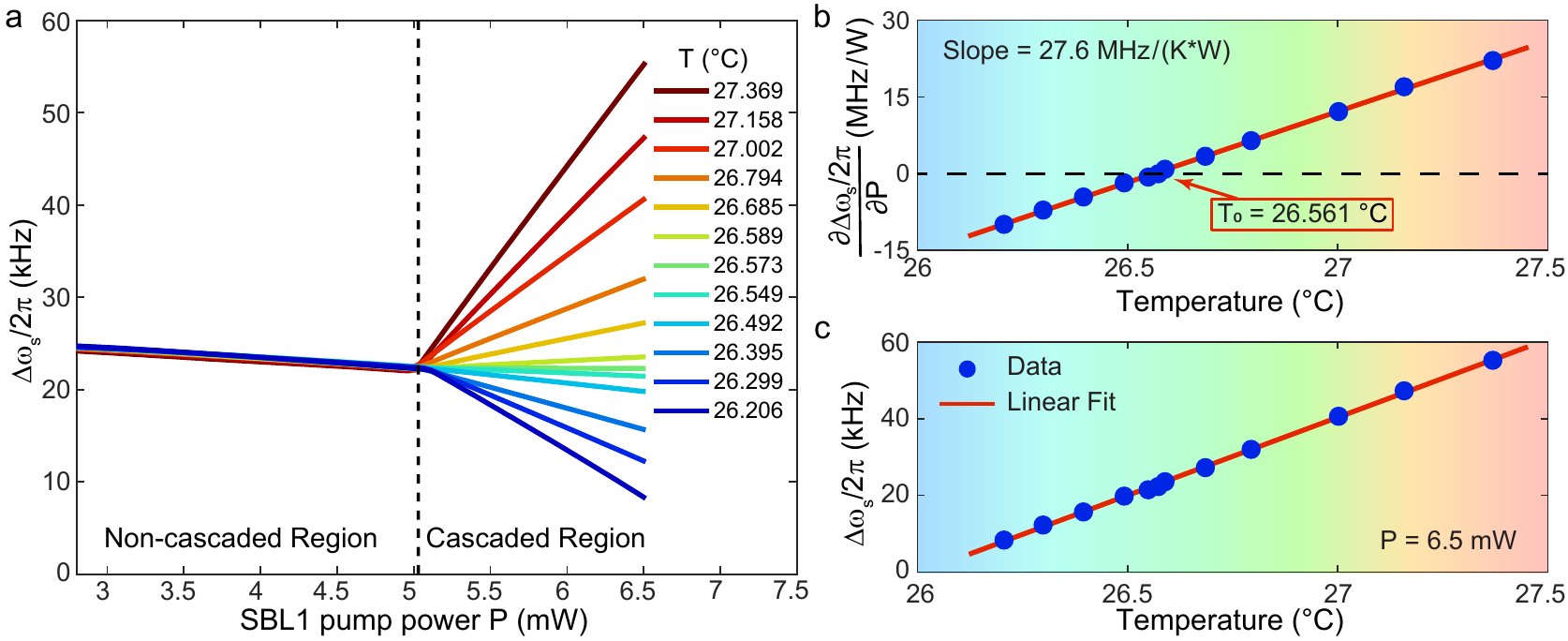}
    \caption{\textbf{Measurement of Brillouin backaction.}
    \textbf{a,} The beating frequency of the counter-propagating SBLs is plotted versus pump 1 power at a series of resonator temperatures as indicated. Frequencies at all temperatures closely track one another in the noncascaded regime with a slight power dependence is induced by the Kerr effect. In the cascaded regime, Brillouin backaction resolves the temperature differences with beating frequency showing a distinct dependence upon pump power. The slope of this dependence changes sign at ${\rm T_0}$ (approximately 26.561 $^{\circ}$C).
    \textbf{b,} Measured SBL beating frequency change per unit pump power plotted versus temperature tuning. The temperature at perfect phase matching condition ($T_0$) is indicated.  \textbf{c,} Measured SBL beating frequency change plotted versus temperature. Slope is 41 kHz/K at 6.5 mW pump power.
}
\label{Fig2}
\end{figure*}

The realization of microresonator-based Brillouin lasers \cite{SBS_CaF2,Carmon,lee2012chemically,Loh2016_Brillouin_Rb_Cell,Rakich2017Laser,LiGyro,KYang,SiNx_SBS} has generated interest in their potential application to compact and potentially integrated Brillouin systems\cite{SBS_Review_Integrated_Circuits}. Moreover, high-coherence Brillouin lasers, featuring short-term linewidths below 1 Hz \cite{lee2012chemically,SBS_Ref,Loh2016_Brillouin_Rb_Cell,SiNx_SBS}, have been used for precision measurement and signal generation. This includes microwave synthesis \cite{SBLSynth,li2014electro}, interrogation of atomic clocks \cite{Loh2020Atomic_Clock_SBS}, and rotation measurement \cite{LiGyro,Earth_Gyro_CPSBL,SiNx_SBS}; all of which require excellent temperature stability of the laser mode volume\cite{Loh2019Temp_SBS}. Here, we demonstrate a new method for temperature stabilization of Brillouin lasers based on backaction produced by the Brillouin anti-Stokes process. This process is shown to provide for phase sensitive locking to a temperature set point $T_0$ given by the following condition, which expresses the Brillouin phase matching condition (see Fig. \ref{Fig1}) \cite{lee2012chemically, SBS_Ref}:
\begin{equation}
    \Omega_B(T_0) = m \times {\rm FSR}(T_0) 
\label{eqn0}
\end{equation}
where $\Omega_B(T)$ is the Brillouin shift at temperature $T$ (see Supplementary Information) and $m \times{\rm FSR}(T)$ is an integer multiple ($m$) of the resonator free-spectral-range (${\rm FSR}(T)$) at temperature $T$. The actual temperature $T_0$ can be set by micro-fabrication control of $FSR$ at a specified operating wavelength.

The short term frequency stability of a stimulated Brillouin laser (SBL) is set by fundamental noise associated with the thermal occupancy of photons involved in the Brillouin process \cite{SBS_Ref}. In high-Q microresonators, this adds a white noise contribution to frequency noise with an equivalent short time linewidth less than 1 Hz \cite{SBS_Ref,LiGyro,SiNx_SBS}. However, on longer time scales the frequency stability is most often set by temperature variations. Here, thermo-optical and thermo-mechanical effects change the cavity resonant frequency \cite{Matsko2007_WGMR_freq_ref_I, Savchenkov2007_WGMR_freq_ref_II,SBS_gain_BW}, while the temperature dependence of the sound velocity causes drifts in the Brillouin frequency shift \cite{silica_sound}. The latter couples temperature to the lasing frequency through mode pulling \cite{SBS_Ref} and can also induce short term linewidth variation through the Brillouin $\alpha$ parameter \cite{Yuan2020Linewidth_enhancement_SBS}. To compensate temperature drift, measurement of cavity temperature using modes belonging to different polarizations was demonstrated \cite{Strekalov2011Temp_birefringent,Fescenko2012Temp_Dual_Mode} and has been recently employed in fiber optic Brillouin laser systems and silicon-nitride chip-resonator systems to stabilize frequency \cite{Loh2019Temp_SBS, Loh2020Atomic_Clock_SBS,UCSB_thermometry}. These dual-polarization modes feature different frequency tuning rates versus temperature, thereby providing a way to convert change in modal temperature to measurement of a frequency change. In contrast to this method which relies upon an external frequency reference to establish locking, the backaction method described here features an intrinsic reference temperature $T_0$ given by Eq. (\ref{eqn0}). It's sensitivity limit is also determined by the fundamental white frequency noise of the laser as opposed to the integrated laser linewidth.

\begin{figure*}
    \centering
    \includegraphics[width=17cm]{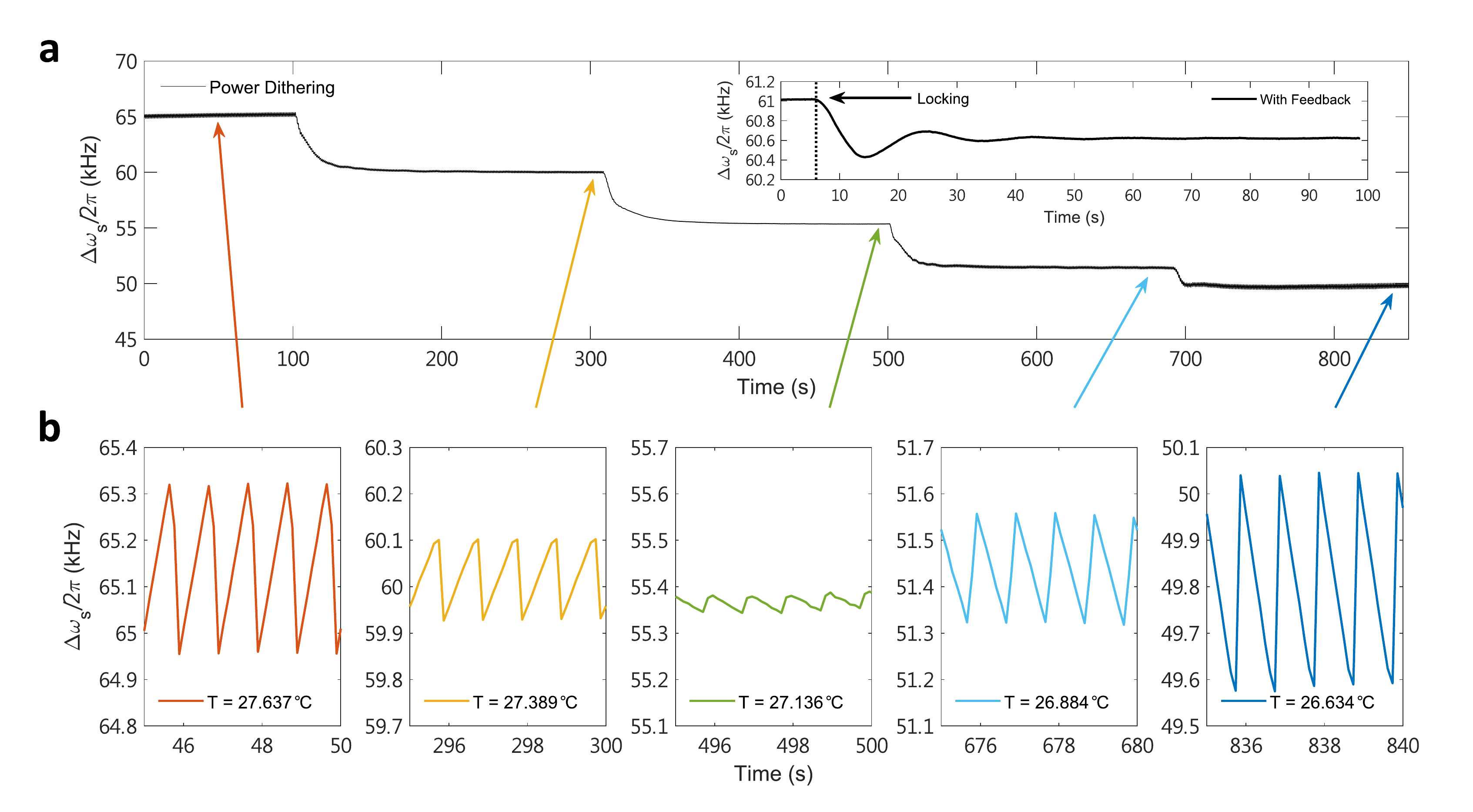}
    \caption{\textbf{Thermal tuning of the backaction with power dithering.}
    \textbf{a,} The open-loop SBL beat frequency (backaction regime of Fig. \ref{Fig2}) is measured versus time at a series of chip temperatures. A weak sawtooth power modulation is applied to Pump 1.
    \textbf{b,} Zoom-ins of the plot in \textbf{a} showing how the polarity and amplitude of the sawtooth beat frequency modulation depends on temperature. \textbf{Inset,} Temperature stabilization of the resonator with servo control activated.}
    \label{Fig3}
\end{figure*}

We consider a Brillouin ring laser geometry shown in Fig. \ref{Fig1}a wherein two pumping waves (dark and light grey arrows) are coupled from a waveguide into clockwise and counter-clockwise directions of the resonator. The frequencies of the pumping waves are close to a resonance, but are not necessarily on resonance. Exact details of this geometry and explanations of the Brillouin process are provided in reference \cite{lai2019observation}. Briefly, each pump wave provides power that is sufficient to excite corresponding Stokes laser waves (SBL1 and SBL2 shown as green and blue arrows) that propagate opposite to their pumping wave and with a lower (Brillouin-shifted) frequency as a result of the phase matching condition. Power transfer from the pumping waves occurs by way of the Stokes scattering process. After laser action occurs, however, a strong anti-Stokes process occurs that is driven by the lasing fields. This creates absorption near the pumps which is proportional to laser power, and this absorptive backaction clamps their circulating powers and hence the laser gain. The exact lasing frequencies can be controlled by tuning of the pumping frequencies, which causes frequency pulling of the laser frequencies such that about 1 MHz of pump frequency tuning induces 10s of kHz of Brillouin laser frequency pulling ($\Delta \omega_s / 2 \pi$). This configuration (non-degenerate, counter-propagating SBLs) is the starting point for implementation of the back-action temperature control method and a schematic of the key spectral features is provided in Fig. \ref{Fig1}b.

Now suppose that the power of SBL1 is increased so that it begins to function as a pumping wave for a new laser wave (the cascaded laser wave shown in red in Fig. \ref{Fig1}b). The power of SBL2 is intentionally kept below threshold so that no cascading occurs. As noted above for the original pumping waves, the onset of laser action, now in the cascaded wave, induces absorptive backaction on its pump, SBL1, that is proportional to the cascaded laser power. It is important to note that on account of the phase matching condition, this backaction acts only SBL1 (i.e., it is directional and does not affect SBL2). The absorptive backaction experienced by SBL1 is shown in purple in Fig. \ref{Fig1}b and compensates SBL1 optical gain (not shown in the figure) provided by the original Pump 1. As a result, SBL1 power is held constant, which is equivalent to the gain of the cascaded laser field (shown in orange in Fig. \ref{Fig1}b) being clamped (even if Pump 1 power is increased). The backaction absorption spectrum has a similar spectral profile to the Brillouin gain and the location of its maximum relative to the frequency of SBL 1 is determined by the phase matching condition for the backaction process. Specifically, maximum nonlinear absorption occurs for the condition of perfect phase matching given by Eq. (\ref{eqn0}).  

The use of this backaction for temperature control is now briefly summarized, but a detailed analysis is given in Supplementary Information. Associated with the backaction absorption is a dispersion contribution that pushes the frequency of SBL1 away from the absorption center. At $T=T_0$ the pushing is zero since this temperature corresponds to perfect phase matching for which the SBL 1 frequency is at the absorption maximum. However, cases where $T<T_0$ and $T>T_0$ (upper and lower panels in Fig. \ref{Fig1}b) result in slight phase mismatch. Defining a phase mismatch parameter as $\Delta \omega \equiv m \times{\rm FSR} - \Omega_B(T)$, for $\Delta \omega > 0$ (upper panel) and $\Delta \omega < 0$ (lower panel) the SBL 1 frequency is pushed higher and lower, respectively. Since the frequency of SBL2 is not affected by this process, measurement of the frequency difference of SBL1 and SBL2 gives a direct way to measure $T-T_0$. As an aside, in reaching this conclusion, it is important to note that cavity resonant frequency tuning with respect to temperature is common mode for SBL1 and SBL2 as they share the same cavity longitudinal mode. 

To implement this measurement it is convenient to use the power dependence of the frequency pushing. Specifically, a weak modulation of the pump 1 power ($P$) will induce a corresponding frequency modulation of SBL1 through modulation of the backaction dispersion. The resulting pump power ($P$) dependence of the SBL2-SBL1 beat frequency ($\Delta\omega_s$) is given by the following equation (see SI):
\begin{equation}
    \frac{\partial \Delta\omega_s}{\partial P} \approx \frac{4g_0\gamma_{ex}}{\hbar\omega_p\gamma^2\Gamma}\frac{d\Omega_B}{dT} (T-T_0)
\label{eqn1}
\end{equation}
where $\Delta\omega_{s}\equiv\omega_{r}-\omega_{s}$, $\omega_{s}$ ($\omega_{r}$) is the absolute angular frequency of SBL1 (SBL2), $g_0$ is the Brillouin gain, $\omega_p$ is the pump angular frequency, $\gamma_{ex}$ is the external coupling rate, $\gamma$ is the total cavity loss rate, and $\Gamma$ is the Brillouin gain bandwidth. This result shows that frequency discrimination of $\Delta\omega_{s}$ combined with subsequent phase sensitive detection will provide an error signal whose magnitude and sign vary as $T-T_0$.

In the experiment, we used a 36 mm-diameter silica wedge resonator on silicon \cite{lee2012chemically} with 8 $\mu$m thickness and wedge angle of 30 degrees. The resonator is packaged (similar to ref. \cite{Earth_Gyro_CPSBL}) with a thermal electrical cooler (TEC), a light emitting diode, and a thermistor.  The TEC and thermistor are used for coarse temperature control and monitoring. The ultra-high-quality factor of the microcavity and the precisely controlled resonator size enable efficient generation of stimulated Brillouin laser action in the opposite propagation directions (operating wavelengths close to $\lambda\approx1553.3$ nm). The intrinsic Q factor is 300 million and the SBL threshold is 0.9 mW. Details on the optical pumping as well as generation of SBL1, SBL2 and the cascaded laser wave are provided in the Figure \ref{Fig1}a caption. 

To setup the phase-sensitive servo temperature control, the beat frequency of SBL1 and SBL2 is dithered by modulating pump 1 (and in turn the cascaded SBL power) using a power modulator (orange dashed line). The frequency of the beat signal is monitored using a frequency counter. A frequency tracking circuit demodulates the dithered signal and phase sensitive measurement is performed using a lock-in amplifier (Stanford Research SR830) to generate the error signal. Temperature control applies the feedback signal to a light emitting diode (fine control) and a thermal electrical cooler (coarse control).

Figure \ref{Fig2} shows the measured SBL1/SBL2 beating frequency ($\Delta \omega_s / 2 \pi$) when sweeping the pump power 1 from below to above cascaded laser operation. Sweeps are performed at a series of temperatures as indicated. As expected, when the pump power is low in the non-cascaded regime, all temperatures provide identical traces. The observed slope on all of these traces is the result of frequency shift provided by the Kerr effect \cite{wang2019petermann}. On the other hand, for higher pump powers in the cascaded regime, the back-action effect is apparent with each temperature showing a distinctly different linear dependence on power. Significantly, the slope of this dependence is observed to change sign as discussed above, corresponding to temperatures above and below $T_0$ in Eq. (\ref{eqn1}). This sign change is essential for implementation of the phase sensitive detection servo control. An experimental plot showing the sign change in slope is provided in Fig. \ref{Fig2}b, where the absolute temperature reference T$_0$=26.561 $^{\circ}$C is also measured. The beating frequency is observed to show a linear power dependence on temperature over the narrow range measured (see Fig. \ref{Fig2}c measured at 6.5 mW pump power). The temperature tuning rate is smaller than the corresponding value provided by the dual-polarization approach \cite{Loh2019Temp_SBS}. Nonetheless, it should be possible to substantially increase this rate in resonators designed for forward-Brillouin scattering, wherein the reduced Brillouin-shift is accompanied by much narrower linewidths (see SI). By fitting these data sets, the experimental back-action strength $\partial^2\Delta\omega_s/(\partial P\partial T)$ is measured to be $2\pi\times27.6$ MHz/(W$\cdot$K), and is consistent with the theory from Eq. (\ref{eqn1}) ($2\pi\times22.5$ MHz/(W$\cdot$K), see SI).

The linear power and temperature dependence of the backaction are further illustrated in Figure \ref{Fig3}a where the beating frequency is measured versus time at a series of temperatures.  To illustrate the change in slope with power at each temperature a weak and slowly-varying saw-tooth power modulation is applied. Zoom-in views of the corresponding saw-tooth modulation in the beat frequency are presented in Fig. \ref{Fig3}b. The change in polarity and amplitude of the backaction-induced modulation are apparent as the temperature is set to values above and below T$_0$.

Finally, long term temperature stabilization of the system is demonstrated by closing the servo control loop. Here, a faster sinusoidal power modulation (200 to 500 Hz) is used to generate a small frequency dither on the SBL beating frequency. It is demodulated by a frequency to voltage conversion circuit and the error signal is generated by phase-sensitive detection with a lock-in amplifier as before. The output of a proportional-integral (PI) servo drives a 1 W white LED (see Figure \ref{Fig1}a) for faster fine-control of temperature. The temperature is also controlled by a TEC that provides slower feedback. The temperature feedback result is shown in the inset of Fig. \ref{Fig3}a. After an initial relaxation oscillation, the long term temperature drift ($>10$ seconds) is stabilized.

\begin{figure}[t]
    \centering
    \includegraphics[width=8.5cm]{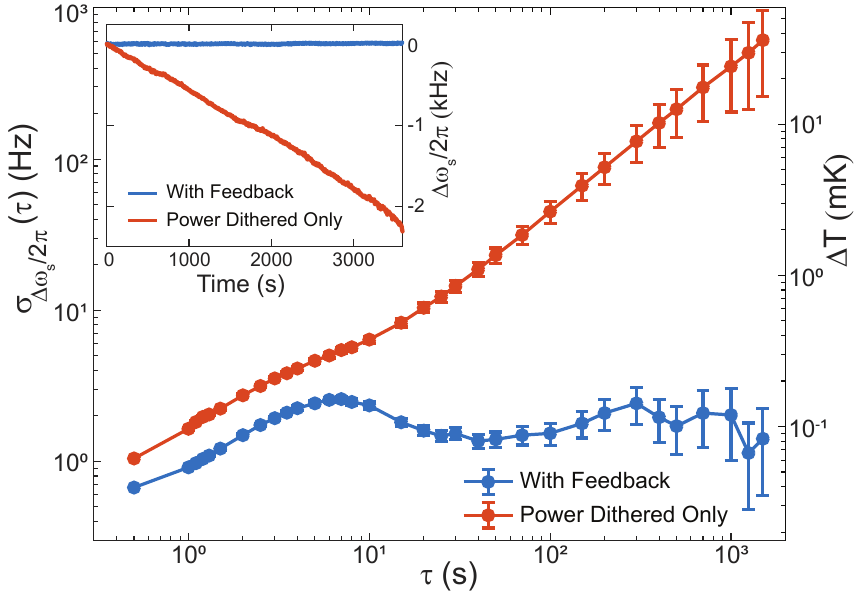}
    \caption{\textbf{Allan Deviation of SBL beating frequency $\Delta\omega_s$.}
    Allan deviations of SBL beating frequency with (blue) and without (red) servo control. The error bar gives the standard deviation.
    \textbf{Inset,} SBL beating frequency versus time with (blue) and without (red) servo control.
}\label{Fig4}
\end{figure}

With the servo-control loop disconnected, but with the power dither active, the beating frequency exhibits a continuous drift as large as 2.3 kHz in an hour, corresponding to around 0.13 $^\circ$C temperature change per hour. This is apparent in both the measured SBL beat frequency (see Fig. \ref{Fig4} inset) and its Allan Deviation (ADEV) measurement presented in the main panel in Fig. \ref{Fig4}). However, with the servo-control loop connected there is no observable drift in the beating frequency over an hour of measurement (see Fig. \ref{Fig4} inset). Also, over 1500 s averaging time (limited by data size of 1 hour) the Allan Deviation remains around 2 Hz, corresponding to about 0.1 mK temperature variation. In the short term, the drift suppression is believed to be limited by the thermal response of the cavity to the LED. In the future, a faster form of temperature feedback (e.g., an integrated resistive heater placed near the resonator) should further reduce this response time. Feed forward frequency correction could also be employed. 

In summary, we have investigated Brillouin backaction and shown that it provides a way to phase sensitively lock an optical resonator to an absolute temperature defined by the phase matching condition. The back-action has been shown to induce both linear power and temperature dependences in a readily measured optical beat frequency. The polarity of the power dependence depends upon operation above or below the phase matching temperature.  This feature and the high stability of the beat frequency were used to servo control the optical mode temperature to 0.1 mK stability levels. While not yet at the temperature stability level of the cross polarization method when applied to microcavities (0.008 mK) \cite{OEwavesTstability}, cross-polarization stabilization has existed for a decade. And we believe the initial results presented here can be substantially improved. An important aspect of the phase sensitive control is that the noise limit is determined by SBL frequency noise at large offset frequencies set by backaction modulation. Here, the rate is 500 Hz, but could be set even higher if necessary to avoid laser frequency noise. Moreover, characterization of the absolute optical frequency stability that is achievable by this method is a possible area of investigation. The range of material systems on which Brillouin laser action has been demonstrated (including integrable platforms) suggests that this method can find wide use. 

\vbox{}
\noindent \textbf{\large Acknowledgments} \\  
This project was supported by the Defense Advanced Research Projects Agency (DARPA) through SPAWAR (grant no. N66001-16-1-4046), the Air Force Office of Scientific Research (FA9550-18-1-0353) and the Kavli Nanoscience Institute. Yu-Kun Lu would like to thank the Caltech SURF program for financial support. 

\vbox{}
\noindent \textbf{\large Author Contributions} \\
Y.H.L. and K.V. conceived the microresonator cascaded Brillouin thermometer. Y.H.L., Y.K.L., Z.Y., H. W. and K.V. constructed the theoretical model. M.G.S. fabricated the ultra-high-Q silica microresonator. Y.H.L., Z.Y., and M.G.S. performed the experiment and analyzed the data. All authors contributed to writing the manuscript. K.V. supervised the project.

\vbox{}
\noindent \textbf{\large Competing interests} \\ 
The authors declare no competing interests.

\vbox{}
\noindent \textbf{\large Data Availability} \\ The data that support the plots within this paper and other findings of this study are available from the corresponding authors upon reasonable request.


\clearpage
\onecolumngrid
\appendix
\renewcommand{\theequation}{S\arabic{equation}}
\renewcommand{\thefigure}{S\arabic{figure}}
\setcounter{figure}{0}
\setcounter{equation}{0}

\section*{Supplementary Information}

Below we derive the temperature dependence of the frequency shift in the power-dithered cascaded SBL. In the cavity-mode rotating frame, we write the pump, SBL, and cascaded SBL in the following form,
\begin{eqnarray}
\dot{A} & = & \left[i\left(\omega_p-\omega_0\right)-\frac{\gamma}{2}\right]A-g_s^*\left|\alpha\right|^2A+\sqrt{\frac{\gamma_{ex}P}{\hbar\omega_p}}, \label{EOM_A1_new} \\
\dot{\alpha} & = & \left[i\left(\omega_s-\omega_1\right)-\frac{\gamma}{2}\right]\alpha+g_s\left|A\right|^2\alpha-g_c^*\left|\beta\right|^2\alpha, \label{EOM_a_new} \\
\dot{\beta} & = & \left[i\left(\omega_c-\omega_2\right)-\frac{\gamma}{2}\right]\beta+g_c\left|\alpha\right|^2\beta, \label{EOM_b_new}
\end{eqnarray}
where $A$, $\alpha$, $\beta$ are the normalized photon number amplitudes of the pump, SBL, and cascaded SBL, respectively.
Here we have adiabatically eliminated the phonon fields as described elsewhere \cite{SBS_Ref}. We also note that the term involving $g^*_c$ in Eq. \ref{EOM_a_new} results from anti-Stokes scattering of the cascaded laser field $\beta$.  The $\omega_p$, $\omega_s$, and $\omega_c$ are the lasing angular frequencies, and the $\omega_0$, $\omega_1$, $\omega_2$ are the cavity angular frequencies. $P$ is the input pump power (i.e., pump 1 in Fig. 1 in the main text). $g_{s,c}$ are defined by,
\begin{eqnarray}
g_{s,c} & = & \frac{g_0}{1+\frac{2i\Delta\Omega_{s,c}}{\Gamma}}, \\
\Delta \Omega_s & = & \omega_p-\omega_s-\Omega_s, \\
\Delta \Omega_c & = & \omega_s-\omega_c-\Omega_c, \\
\Omega_B & \equiv & \Omega_s \approx \Omega_c,
\end{eqnarray}
where $\Omega_s$ and $\Omega_c$ are the phonon angular frequencies associated with the SBL and the cascaded SBL, respectively. $\Omega_B$ is the Brillouin shift, which is equal to $4\pi n c_s/\lambda_p$ ($n$ is the refractive index, $c_s$ is the speed of sound in silica, and $\lambda_p$ is the pump wavelength).  $g_0$ is the Brillouin nonlinear coefficient \cite{SBS_Ref}, $\gamma_{ex}$ is the external coupling coefficient, and $P$ is the input pump power.

In steady-state, if we assume good phase matching ($\Delta\Omega_{s,c}\ll\Gamma$), then the real parts and imaginary parts of Eq.(\ref{EOM_a_new}) and Eq.(\ref{EOM_b_new}) give,
\begin{eqnarray}
\frac{\gamma}{2} & = & g_0\left(\left|A\right|^2-\left|\beta\right|^2\right) = g_0\left|\alpha\right|^2, \label{real_g0_new} \\
\omega_s-\omega_1 & = & \frac{2g_0}{\Gamma}\left(\left|A\right|^2\Delta\Omega_s+\left|\beta\right|^2\Delta\Omega_c\right) \label{omega_s_before_new} \\
\omega_c-\omega_2 & = & \frac{2g_0}{\Gamma}\left|\alpha\right|^2\Delta\Omega_c. \label{omega_c_before_new}
\end{eqnarray}
From Eq. (\ref{real_g0_new}) we get a photon number relation $|A|^2=|\alpha|^2+|\beta|^2$ as well as the clamping condition for the Stokes wave, $|\alpha|^2=2g_0/\gamma$. These can be used to eliminate $|\alpha|^2$ and $|\beta|^2$ in Eq. (\ref{omega_s_before_new}) and Eq. (\ref{omega_c_before_new}), which yields,
\begin{eqnarray}
\omega_s - \omega_1 & = & \frac{2g_0}{\Gamma}\left|A\right|^2\left(\Delta\Omega_s+\Delta\Omega_c\right)-\frac{\gamma}{\Gamma}\Delta\Omega_c, \label{omega_s_new} \\
\omega_c - \omega_2 & = & \frac{\gamma}{\Gamma}\Delta\Omega_c. \label{omega_c_new}
\end{eqnarray}
To study the dependence of $\omega_s$ on the input power, we take the partial derivative with respect to $|A|^2$. Since $\omega_1$, $\omega_2$, $\omega_p$, $\Omega_B$ are independent of power the following hold,
\begin{eqnarray}
\frac{\partial \omega_c}{\partial\left|A\right|^2} & = & \frac{\gamma}{\Gamma}\frac{\partial \Delta\Omega_c}{\partial\left|A\right|^2} =\frac{\gamma/\Gamma}{1+\gamma/\Gamma}\frac{\partial\omega_s}{\partial\left|A\right|^2} \label{omega_c_P_new} \\
\frac{\partial\omega_s}{\partial\left|A\right|^2} & = & \frac{2g_0}{\Gamma}\left(\omega_p-\omega_c-2\Omega_B\right) \nonumber \\
&+& \frac{2g_0}{\Gamma}\left|A\right|^2\frac{\partial \left(\Delta\Omega_s+\Delta\Omega_c\right)}{\partial \left|A\right|^2}-\frac{\gamma}{\Gamma}\frac{\partial \Delta\Omega_c}{\partial \left|A\right|^2} \nonumber \\
&=&  \frac{2g_0}{\Gamma}\frac{\omega_p-\omega_c-2\Omega_B}{1+\frac{\gamma/\Gamma}{1+\gamma/\Gamma}\left(1+\frac{2g_0\left|A\right|^2}{\Gamma}\right).}
\label{omega_s_P_new}
\end{eqnarray}

Eq. (\ref{omega_c_P_new}) shows that the cascaded Brillouin laser frequency is also affected by the mode pulling effect\cite{lai2019observation}, such that the cascaded laser frequency moves toward to the cavity mode center and is less sensitive to the original Brillouin laser drift. Assuming the operation is only slightly above the cascaded threshold, such that Eq. (\ref{real_g0_new}) gives $2g_0\left|A\right|^2\approx\gamma$, then simplifies Eq. (\ref{omega_s_P_new}) into
\begin{eqnarray}
\frac{\partial \omega_s}{\partial \left|A\right|^2} & = & \frac{1}{1+\gamma/\Gamma}\frac{2g_0}{\Gamma}\left(\omega_p-\omega_c-2\Omega_B\right).
\label{omega_s_P_new2}
\end{eqnarray}

Then, by assuming $\gamma/\Gamma\ll 1$, we can drop the correction factor in Eq. (\ref{omega_s_P_new2}). Next, the $\left|A\right|^2$ can be further replaced by $P = \hbar\omega_p\gamma^2\left|A\right|^2/\gamma_{ex}$, which is the input pump power above cascade threshold (obtained by solving the steady state of Eq. (\ref{EOM_A1_new}) with clamping conditions for $\left|\alpha\right|^2$). Then, Eq. (\ref{omega_s_P_new2}) simplifies to the following,
\begin{eqnarray}
\frac{\partial \omega_s}{\partial P} & = & \frac{2g_0\gamma_{ex}}{\hbar\omega_p\gamma^2\Gamma}\left(\omega_p-\omega_c-2\Omega_B\right).
\label{omega_s_P}
\end{eqnarray}
The derivative with respect to temperature is now taken in this expression to arrive at the following,
\begin{eqnarray}
\frac{\partial^2\omega_s}{\partial P\partial T} & \approx & -\frac{4g_0\gamma_{ex}}{\hbar\omega_p\gamma^2\Gamma}\frac{d\Omega_B}{dT}
\label{Temp}
\end{eqnarray}
In writing this expression the temperature dependence of 
$\omega_p-\omega_c\approx\omega_0-\omega_2\approx 2m\times {\rm FSR}$ is neglected. This term's temperature dependence is dominated by the thermorefractive and thermoexpansion effects, which are much weaker than the temperature dependence of the Brillouin frequency. For example, in silica glass the thermoexpansion coefficient is $\alpha_L = 0.51\times10^{-6}/$K and the thermorefractive coefficient is $dn/dT=11.6\times10^{-6}/$K, so that $d\left(\omega_p-\omega_c\right)/dT \approx 2\pi\times 180$ kHz/K. This compares to $d\Omega_B/dT\approx2\pi\times 1.16$ MHz/K at $1550$ nm (estimated from $1.36$ MHz/K at $1320$ nm \cite{SBS_gain_BW}), and justifies the simplification involved in Eq. (\ref{Temp}). 

Now, if we introduce an independent backward propagating SBL as a reference with higher angular frequency $\omega_r$, the result is,
\begin{equation}
    \frac{\partial^2\Delta\omega_s}{\partial P\partial T} = \frac{\partial^2(\omega_r-\omega_s)}{\partial P\partial T} = \frac{4g_0\gamma_{ex}}{\hbar\omega_p\gamma^2\Gamma}\frac{d\Omega_B}{dT}
\label{Temp2}
\end{equation}
which is the temperature derivative of Eq. 2 in the main text. The silica resonator has $g_0/2\pi=0.61$ mHz, $\Gamma/2\pi=30$ MHz, $\gamma_{ex}/2\pi = 110$ kHz, $\gamma/2\pi=860$ kHz, $\omega_p/2\pi=193$ THz, giving a theoretical estimation of $\partial^2\Delta\omega_s/\partial P\partial T=2\pi\times22.5$ MHz/(W.K), which compares favorably with the experimental value in Fig. \ref{Fig2}) of $2\pi\times27.6$ MHz/(W.K). The difference here mainly originates from the uncertainty of certain parameters.

As an aside, Eqs. (\ref{Temp}) and (\ref{Temp2}) show that frequency tuning response with respect to temperature depends inversely upon the Brillouin damping rate $\Gamma$. Along these lines, forward Brillouin scattering in high-Q silica microresonators has produced damping rates much smaller than for back scattering on account of the smaller required Brillouin shift for backscattering \cite{Carmon, Carmon2}. We intend to investigate this approach as a means to increase the backaction response.

To estimate $g_0$ above (in the unit of rad/s), we use the following equations\cite{Cryogenic_SBS}:
\begin{eqnarray}
g_0 & \approx & \frac{\hbar\omega^3}{2P_{\textrm{clamp}}Q_TQ_E} \\
& \approx & \frac{2\pi\Delta\nu_{\textrm{clamp}}}{n_{th}},
\end{eqnarray}
where $P_{\textrm{clamp}}$ is the clamped power of SBL at the cascading threshold, $Q_T$ ($Q_E$) is the total (external) quality factor, $\Delta\nu_{\textrm{clamp}}$ the full-width-half-maximum of the fundamental SBL linewidth under cascaded clamping conditions, $n_{th}$ is the number of thermal phonon quanta at the operating temperature. For the 36 mm silica resonator at room temperature, we measured $\Delta\nu_{\textrm{clamp}}=0.35$ Hz and used the theoretical $n_T\approx577$.

\end{document}